# CellProfiler Analyst 3.0: Accessible data exploration and machine learning for image analysis.

David R. Stirling[1], Anne E. Carpenter[1] and Beth A. Cimini[1*]

[1]=Imaging Platform, Broad Institute of MIT and Harvard, Cambridge MA 02142

*= To whom correspondence should be addressed

**Abstract**

Image-based experiments can yield many thousands of individual measurements describing each object of interest, such as cells in microscopy screens. CellProfiler Analyst is a free, open-source software package designed for the exploration of quantitative image-derived data and the training of machine learning classifiers with an intuitive user interface. We have now released CellProfiler Analyst 3.0, which in addition to enhanced performance adds support for neural network classifiers, identifying rare object subsets, and direct transfer of objects of interest from visualisation tools into the Classifier tool for use as training data. This release also increases interoperability with the recently released CellProfiler 4, making it easier for users to detect and measure particular classes of objects in their analyses.

**Background**

With the increasing adoption of high-throughput microscopy, scientists have been able to generate large datasets containing thousands of individual images. This necessitates automated computational analysis to efficiently derive biological insights from the raw data. Free software packages such as ImageJ [1] and CellProfiler [2] allow users to extract hundreds or thousands of numerical measurements from their image data, but it is ultimately on the user to determine which of these features are relevant to the biological problem being investigated. Spreadsheet programs are familiar but lack features and integration with source images that biologists often need. CellProfiler Analyst is a data exploration package for helping users to explore and extract information from large datasets, including (but not limited to) those produced by CellProfiler pipelines [3]. The software includes several tools for users to visualise and filter their datasets, alongside tools for training machine learning classifier models within a convenient graphical user interface that is geared towards working with image data [4]. While other tools such as Ilastik [5] and Advanced Cell Classifier [6] can provide a GUI for training classifiers with image-based data, these do not include data exploration and visualisation tools like those in CellProfiler Analyst. Herein we present CellProfiler Analyst 3.0, which includes major performance improvements and new features that improve the utility of the software.

**General Changes**

We ported CellProfiler Analyst to the Python 3 programming language to ensure compatibility with future operating systems after the official Python 2 end-of-life in 2020. We also revised the program's builds to package all Java dependencies within the main installer, which dramatically simplifies the installation process. In addition, we added a faster imageio-based image loader to supplement the existing bioformats-based implementation [7]. While bioformats provides broader file format compatibility, imageio allows for common image formats to be loaded more efficiently, which greatly improves the time to generate object thumbnails.

CellProfiler Analyst 3.0 can export machine learning models that are compatible with CellProfiler 4.2+, allowing the resulting classifiers to be directly embedded into CellProfiler pipelines. This maintains and expands upon the interoperability between the two programs.

**New Features**

We added a dimensionality reduction tool to help users visualise the variance within high-dimensional data sets (Figure 1A). This is particularly important when creating classifiers that can identify rare images in a set; rare outliers are needed for training but may be difficult to find through random sampling. When the data set contains thousands of measurements users can struggle to identify those which reveal these outliers. Dimensionality reduction allows the numerous measurements generated by CellProfiler to be condensed into a smaller subset of features which represent the overall variance in the data set. This provides a more manageable series of features which the user can then explore.

This tool supports multiple common reduction techniques including Principal Component Analysis (PCA) [8], Feature Agglomeration [9] and t-Distributed Stochastic Neighbor Embedding (t-SNE) [10]. The resulting components can be visualised on a scatter plot, which can color individual objects by class if classification was previously performed on the data set. As with other tools, users can create gates to isolate specific populations for further analysis.

We improved the gating functionality so that gates drawn in plotting tools are immediately made available within the Classifier tool, rather than having to be manually converted into filters. The dimensionality reduction tool contains a new lasso tool for selecting objects of interest through drawing a custom polygon. Right-click menu options allow the selected objects to be sent directly to an open Classifier window for use as training data. These improvements simplify the process of finding populations of interest and creating training sets for machine learning models in the Classifier.

In the Classifier tool we added support for automatic rescaling of input data, based on the scikit-learn StandardScaler implementation [11]. This improves performance of certain classifier types (such as K-Neighbors) that are skewed by the absolute values of measurements. It also now supports tunable neural network classifiers based on the scikit-learn MLPClassifier class [12], which can be useful for performing complex non-linear classification tasks. New keyboard shortcuts for sorting objects into classification bins provide a more rapid means of generating training sets than dragging-and-dropping.

**Performance Improvements**

We improved performance in several areas of the Classifier tool. We optimised loading of previously saved training sets by batching objects during database fetching, reducing the loading time for a sample dataset by over 10-fold (Figure 1B). We also found that removing redundant database calls and unnecessary caching steps during classifier model training reduced processing time by 90% (Figure 1C).

Database handling improvements were also made in the functions for scoring the data sets after training. These refinements produced a modest improvement in performance of the 'score all' (Figure 1D) function. We further optimised the 'score image' function by revising the workflow for fetching object coordinates, which significantly reduced the time taken to display a scoring preview overlay (Figure 1E). Usability improvements to the image viewer include a shortcut to resize the image to fit the window and a table display outlining counts of each class found in the image.

**Future Directions**

The next major frontier for phenotype classification is to train deep learning models straight from raw pixels rather than requiring a separate step feature extraction [13]. Enabling scoring of phenotypes that fall along a continuum rather than into discrete bins would also be useful, as we recently found in red blood cell aging [14]. Ultimately, maintaining CellProfiler Analyst as an up-to-date resource for users to explore high-dimensional, image-based data without needing to code will help the biology community for years to come.

**Availability and Implementation**
CellProfiler Analyst binaries for Windows and MacOS are freely available for download at https://cellprofileranalyst.org/. Source code is implemented in Python 3 and is available at https://github.com/CellProfiler/CellProfiler-Analyst/. A sample data set is available at https://cellprofileranalyst.org/examples, based on images freely available from the Broad Bioimage Benchmark Collection (BBBC).

**Acknowledgements**
The authors would like to thank Pearl Ryder, Erin Weisbart, Jane Hung, David Dao, Egor Zindy and Mario Emmenlauer for contributions to bug fixes and testing of pre-release versions of this software. We also thank all the members of the bioimaging community who have provided feedback and suggestions which have helped to guide this work.

**Competing interests**
The authors declare that they have no competing interests.

**Funding**
This work was supported by National Institutes of Health grants (R35 GM122547 and P41 GM135019 to AEC). This project has been made possible in part by grant number 2020-225720 to BAC from the Chan Zuckerberg Initiative DAF, an advised fund of the Silicon Valley Community Foundation. The funders had no role in study design, data collection and analysis, decision to publish, or manuscript preparation.

## Authors' contributions

DRS developed the software. DRS wrote the manuscript, with editorial contribution and supervision from AEC and BAC. All authors have read and approved the manuscript.

## Figure

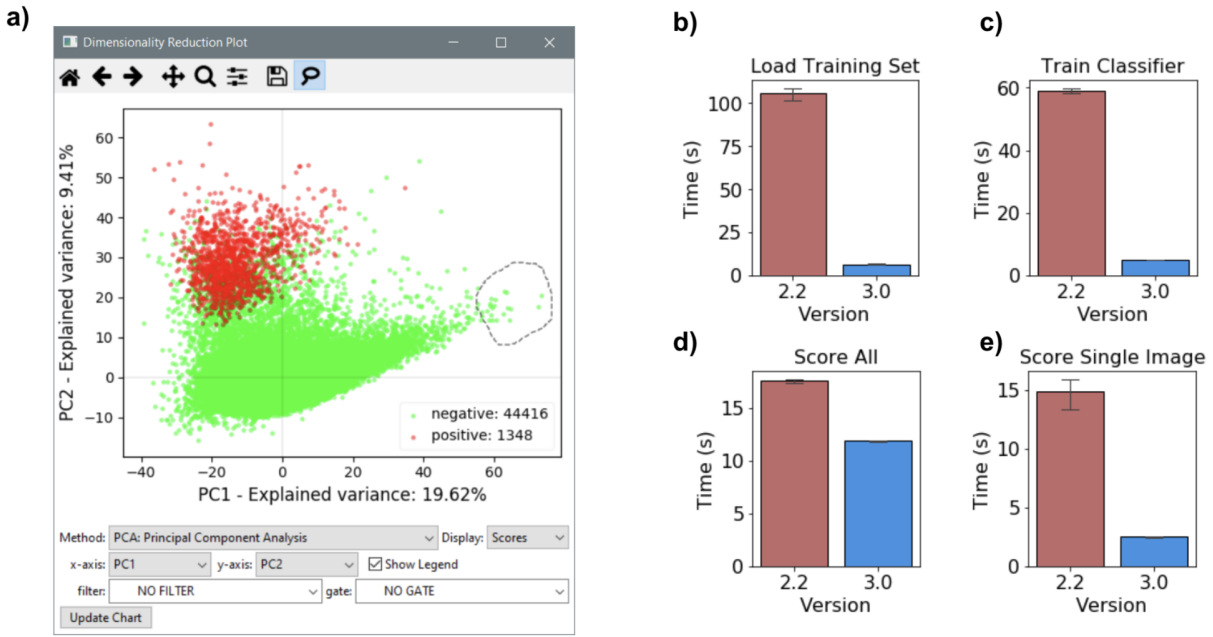

**Figure 1 – Improvements in CellProfiler Analyst 3.0.**
**a)** Screenshot of the new Dimensionality Reduction tool, using principal component analysis with the example dataset. Marker colours represent classification results from a previously generated classifier. The lasso tool has been activated to select objects of interest, which can then be used to train a classifier. **b)** Comparison of execution time between CPA 2.2 and CPA 3.0 when loading a training set of 600 objects in 13 classes. **c)** Execution time to run the training sequence for a FastGentleBoosting classifier. **d)** Time taken to score 45,000 objects using a FastGentleBoosting classifier. **e)** Execution time to score a single image containing 130 objects, then annotate results onto a preview image.